\documentclass[manuscript]{aastex}

\newcommand{\hii}{H\textsc{ii}}
\def\ks{km s$^{-1}$}
\def\deg{$^\circ$}
\def\m{$^\prime$}
\def\s{$^{\prime\prime}$}
\def\hh{$^{\mathrm h}$}
\def\mm{$^{\mathrm m}$}
\def\ss{$^{\mathrm s}$}
\def\cm3{cm$^{-3}$}

\def\2{$^{12}$CO}
\def\3{$^{13}$CO}
\def\H{HCO$^{+}$}
\def\msun{M$_\odot$}

\shorttitle{The HII region G46.5-0.2 and surrounding molecular gas}
\shortauthors{Paron et al.}

\begin{document}


\title{HII region G46.5-0.2: the interplay between ionizing radiation, molecular gas and star formation}



\author{S. Paron\altaffilmark{1,2}, M. E. Ortega\altaffilmark{1}, G. Dubner \altaffilmark{1}, Jing-Hua Yuan \altaffilmark{3},
A. Petriella\altaffilmark{1}, E. Giacani\altaffilmark{1,2}, Jin Zeng Li\altaffilmark{3}, 
Yuefang Wu\altaffilmark{4}, Hongli Liu\altaffilmark{3}, Ya Fang Huang\altaffilmark{3}, and Si-Ju Zhang\altaffilmark{3} }
\email{sparon@iafe.uba.ar}






\altaffiltext{1}{Instituto de Astronom\'{\i}a y F\'{\i}sica del Espacio (IAFE, CONICET-UBA),
             CC 67, Suc. 28, 1428 Buenos Aires, Argentina}
\altaffiltext{2}{FADU - Universidad de Buenos Aires, Ciudad Universitaria, Buenos Aires, Argentina}
\altaffiltext{3}{National Astronomical Observatories, Chinese Academy of Sciences, 20A Datun Road, Chaoyang District, Beijing 100012, China}
\altaffiltext{4}{Department of Astronomy, Peking University, 100871 Beijing, China}

\begin{abstract}
\hii~regions are particularly interesting because they can generate dense layers of gas and dust, elongated columns or pillars
of gas pointing towards the ionizing sources, and cometary globules of dense gas, where triggered star formation can
occur. Understanding the interplay between the ionizing radiation and the dense surrounding gas is very
important to explain the origin of these peculiar structures, and hence to characterize triggered star formation. 
G46.5-0.2 (G46), a poorly studied galactic HII region located at about 4 kpc, is an excellent target to perform this kind of studies.
Using public molecular data extracted from the Galactic Ring Survey ($^{13}$CO J=1--0) and from the James Clerk Maxwell Telescope data 
archive ($^{12}$CO, $^{13}$CO, C$^{18}$O J=3--2, HCO$^{+}$ and HCN J=4--3), and infrared data from the GLIMPSE and MIPSGAL surveys, we 
perform a complete study of G46, its molecular environment and the young stellar objects placed around it.
We found that G46, probably excited by an O7V star, is located close to the edge of the GRSMC G046.34-00.21 molecular cloud. It presents
a horse-shoe morphology opening in direction of the cloud. We observed a filamentary structure in the molecular gas 
likely related to G46 and not considerable molecular emission towards its open border.  We found that about 10$^\prime$~towards the 
southwest of G46 there are
some pillar-like features, shining at 8~$\mu$m and pointing towards the \hii~region open border. We propose that the 
pillar-like features were carved and sculpted by the ionizing flux from G46. We found several young stellar objects likely embedded 
in the molecular cloud grouped in two main concentrations: one, closer to the G46 open border consisting of Class II type sources, 
and other one mostly composed by Class I type YSOs located just ahead the pillars-like features, strongly suggesting an age gradient in the 
YSOs distribution.

\end{abstract}


\section{Introduction}

Massive stars play a key role in the evolution of the Galaxy.
They are the principal source of heavy elements and
UV radiation. Through the combination of winds, massive
outflows, expanding \hii~regions and supernova explosions,
they provide an important source of mixing turbulence in
the interstellar medium (ISM). 
\hii~regions are particularly interesting because they can 
trigger the formation of a new generation of stars. Processes such
`collect and collapse' (e.g. \citealt{elme95}) and radiative driven implosion (RDI) (e.g. \citealt{berto89}) have been 
convincingly demonstrated that occur around \hii~regions leading to formation of stars (e.g. 
\citealt{ortega13,diri12,paron11,zav10,deha08,thompson04a}).
These processes generate dense layers of gas and dust at the interface between
the \hii~regions and their parent molecular cloud, elongated columns (pillars) of gas pointing 
towards the ionizing sources, and cometary globules of dense gas \citep{tremb13}. 
Understanding the interplay between the ionizing radiation and the dense surrounding gas is very  
important to explain the origin of these peculiar structures, and hence to characterize triggered star formation.

G46.5-0.2 (hereafter G46) is a poorly studied \hii~region
about 8\m\, in size located at RA $=$ 19\hh17\mm26\ss, dec $=$ 11\deg55\m54\s~(J2000), 
on the border of the molecular cloud catalogued as GRSMC G046.34-00.21 \citep{rath09}. 
\citet{lockman89} observed a recombination line at v$_{\rm LSR} \sim$ 57.2 \ks~towards this
region. \citet{kuchar94} detected  HI absorption features up to
56.6 \ks~and based on kinematical considerations proposed for this \hii~region
the near distance of 3.8~kpc, while \citet{quireza06}, using
the same techniques, favored the far distance of 7.8~kpc. Finally,
\citet{anderson09} resolved the kinematic distance ambiguity for G46
based on existing HI and $^{13}$CO sky surveys, confirming a
near distance of 4~kpc, which will be adopted in what follows.

\begin{figure}
\includegraphics[width=8.5cm]{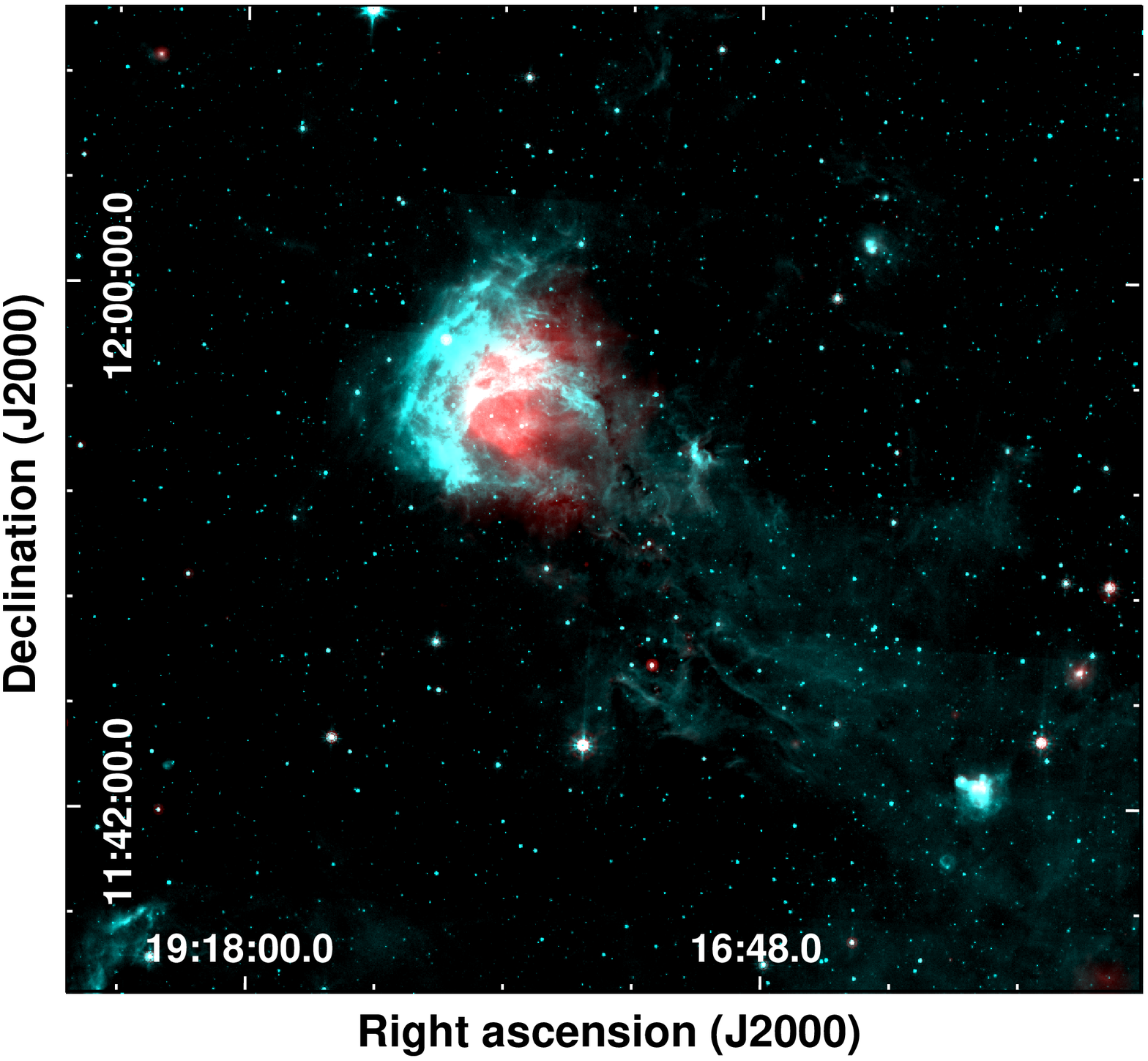}
\caption{Two-color composite image of a large area towards the \hii~region G46. In cyan
the {\it Spitzer}-IRAC 8 $\mu$m emission, and in red the {\it Spitzer}-MIPSGAL emission at 24 $\mu$m .}
\label{present}
\end{figure}

Figure \ref{present} shows a two-color composite image of G46 field 
where the {\it Spitzer}-IRAC 8 $\mu$m emission distribution is displayed in cyan and the {\it Spitzer}-MIPSGAL
at 24 $\mu$m in red. The emission displayed in cyan mainly corresponds to radiation originated in
polycyclic aromatic hydrocarbon (PAH) molecules.  The far-UV photons
leaking from the \hii~region excite the PAHs, which then emit in the
mid-infrared. This emission encircles the 24~$\mu$m emission (in red) 
that arises from very small size dust grains which are heated but not destroyed by the UV
photons. The emission  at 8~$\mu$m has a horse-shoe morphology, being thicker and more
intense towards the east-northeast direction and open to the southwest, where it dilutes into the 
surrounding medium. Precisely in the southwest direction, about 10\m~away from G46 there are pillar-like features shining at 8~$\mu$m. 
In this paper we explore 
the interrelationship among the  \hii~region G46, the environmental conditions and the young stellar objects in the region, 
underscoring evidence of triggered stellar formation in the immediate vicinity of G46 and farther.

\section{Data}

\subsection{Molecular data} 

The \3 J=1--0 data were extracted from the Galactic Ring Survey (GRS). 
The survey, performed by the Boston University and the Five College Radio
Astronomy Observatory, maps the Galactic Ring in
the mentioned molecular line with an angular and spectral resolutions of
46\s~and 0.2 \ks, respectively (see \citealt{jackson06}). The
observations were performed in both position-switching and on-the-fly
mapping modes, achieving an angular sampling of 22\s. Data are presented 
in main beam temperature ($T_{mb}$).

Additionally we used \2, \3, C$^{18}$O J=3--2, and \H, HCN J=4--3 data 
extracted from the James Clerk Maxwell Telescope (JCMT) data archive\footnote{www.jach.hawaii.edu/JCMT/archive/}. 
These observations (Proposal I.D M10AH02, P.I. Jonathan Williams) were carried out with
the JCMT in Mauna Kea, Hawaii, using the HARP-ACSIS instrument. 
The angular and spectral resolutions are about 15\s~and 0.05 \ks. 
We used the reduced data (cal. level 2), which were reduced using the standard ORAC-DR 
pipelines\footnote{http://www.jach.hawaii.edu/JCMT/archive/CADC\_quickguide.html}. 
Data are presented in units of corrected antenna temperature $T_{A}^{*}$, which is related to 
the main beam temperature ($T_{mb}$) using $T_{mb} = T_{A}^{*}/\eta_{mb}$. 
Following \citet{buckle09} a value of $\eta_{mb} = 0.6$ was used.  The zenith opacity was
between 0.04 and 0.06 for all the observed lines.
Even though we extracted reduced 
data, in the case of C$^{18}$O J=3--2, \H, and HCN J=4--3 data we applied a Hanning smoothing to improve the signal-to-noise ratio 
which altered the spectral resolution in no more than a factor of 2, 
and minor polynomials were used 
for baseline corrections using XSpec\footnote{XSpec is a spectral line reduction package for astronomy which has been developed 
by Per Bergman at Onsala Space Observatory.}.  

\subsection{Infrared data}

In this work, archived infrared data are also used to reveal star formation in G46. 
These data come from the GLIMPSE and MIPSGAL surveys.

GLIMPSE \citep[Galactic Legacy Infared Mid-Plane Survey Extraodinaire,][]{ben03} using the IRAC 
\citep[using the InfraRed Array Camera,][]{faz04} on board the {\it Spitzer Space Telescope} \citep{wer04} 
surveyed the inner 130 degrees of the Galactic Plane at 3.6, 4.5, 5.8, and 8.0 \micron~bands. The 5 $\sigma$ 
sensitivities of the four bands are 0.2, 0.2, 0.4, 0.4 mJy, respectively. In addition to images, GLIMPSE survey 
performed point-source photometry. Photometric data at the {\it J} (1.25 $\mu$m), 
{\it H} (1.65 $\mu$m), and {\it Ks} (2.17 $\mu$m) bands from the Two Micron All Sky Point Source Catalog 
\citep[2MASS PSC, ][]{skr06} are provided in the GLIMPSE Catalog to build-up a 7-bands photometric system. Both the images 
and the catalog of GLIMPSE are publicly available at the InfraRed Science Archive (IRSA)\footnote{http://irsa.ipac.caltech.edu/index.html}, 
where we have retrieved cutouts in the four IRAC bands and point sources in a $31\arcmin\times24\arcmin$ region centered at 
$\alpha_{2000}={\rm 19^h17^m18^s.7}$, $\delta_{2000}=+11$\deg$52\arcmin22\arcsec.9$. 
We restricted the extracted catalog to be a more reliable data set with the following criteria: a) only sources with 
photometric errors no larger than 0.2 mag in the IRAC bands are taken into account; b) for 2MASS bands, a threshold of 0.1 mag photometric 
error is used to forsake unreliable photometric values.
The photometric errors for IRAC and 2MASS sources have been adopted following \citet{gut09}.

MIPSGAL \citep{car09} is a complement to the GLIMPSE legacy survey. This survey using the MIPS 
\citep[Multiband Infrared Photometer for {\it Spitzer},][]{rie04} instrument on board the {\it Spitzer Space Telescope} surveyed an 
area comparable to that of GLIMPSE. The version 3.0 of MIPSGAL data includes mosaics only at 24 $\mu$m with sky coverage 
of $|b|<1$\deg~for $-68$\deg$<l<69$\deg, and $|b|<3$\deg~for $-8$\deg$<l<9$\deg. The spatial resolution and $5\sigma$ sensitivity 
at 24 $\mu$m are $6\arcsec$ and 1.7 mJy. From the IRSA server, we have extracted a cutout of a region same as that of GLIMPSE cutouts.

We conducted point-source extraction and aperture photometry of point sources in the MIPS 24 \micron~image using the PSF fitting 
capability of IRAF/DAOPHOT \citep{1987PASP...99..191S}. The PSF was determined to be about 6\arcsec~by fitting the profiles of ten 
bright point sources in the investigated field. The standard deviation ($\sigma$) of the sky was estimated to be about 
$5.0\times10^{-6}~\mathrm{Jy~pixel^{-1}}$. \textsc{daofind} was used to extract candidates of point sources with a threshold of 10 $\sigma$. 
The final sources were determined by visual inspection using the \textsc{tvmark} task. Sources affected by ghosts, diffraction spikes, halos 
from bright sources and artifacts residing in bright extended emission were rejected. For aperture photometry of the extracted sources, 
radii of the apertures and inner and outer limits of the sky annuli were selected to be 4.5\arcsec, 15\arcsec~and 21\arcsec, respectively. 
Magnitudes of the extracted sources were determined using the magnitude zero point of 7.17 Jy provided in the MIPS instrument 
handbook\footnote{http://irsa.ipac.caltech.edu/data/SPITZER/docs/mips/}. Finally, we cross-matched the 24 \micron~sources with the 
GLIMPSE catalog using a cone radius of 2\arcsec.

Additionally, we have queried the MIPSGAL point source catalog given by \citet{gut15}. 
A total of 309 sources are found in the region of interest. Among them, 305 have been covered by 
our 24 $\mu$m point source list. The other four sources are weaker than 11.5 mag. 
A comparison of the fluxes indicates that our photometry is consistent with that given in \citet{gut15}, 
the mean difference is smaller than 5\% in magnitude. 

The completeness of our catalog was estimated by counting the number of sources as a function of magnitude. 
A histogram plot of source magnitudes at each band has been plotted and carefully inspected. 
The sources magnitudes are exponentially distributed. We plot the number of sources in log scale and 
the magnitude in linear scale, then a straight line can be fitted. The magnitude at which a deviation emerges has been 
considered as the completeness limit.
We found that our catalog 
is complete to a magnitude of 14.0 at 3.6 $\mu$m, 13.5 at 4.5 $\mu$m, 12.3 at 5.8 $\mu$m, 12.0 at 8.0 $\mu$m, 
and 8.5 at 24 $\mu$m. For the central region with diffuse mid-infrared emission, our catalog is complete 
to a magnitude of 14.0 at 3.6 $\mu$m, 13.5 at 4.5 $\mu$m, 12.0 at 5.8 $\mu$m, 11.5 at 8.0 $\mu$m, and 8.0 at 24 $\mu$m.

\section{Results and discussion}

In what follows we separately analyze the physical properties of the \hii~region G46 (Sec. 3.1), 
of the ambient molecular gas (Sect. 3.2) and of the young stellar objects (YSOs) near G46 (Sec. 3.3), with the aim 
of investigating the impact of the ionizing radiation on the surrounding medium and the likelihood of triggered star formation.

\subsection{Exciting star(s) candidate(s) and morphology of G46}
\label{star}

A search for the exciting star of G46 in the
available OB-type stars catalogues \citep{maiz13,maiz04,reed03}, was unsuccessful. We can, anyway, conjecture
the earliest spectral type of a probable single ionizing star from the radio continuum emission of the \hii~ 
region. The number of UV ionizing photons needed to keep an \hii~region
ionized is given by N$_{\rm UV} = 0.76 \times 10^{47} T_4^{-0.45} \nu_{\rm
GHz}^{0.1} S_{\rm \nu} D_{\rm kpc}^2$ \citep{cha76}, where $T_4$ is
the electron temperature in units of 10$^4$~K, $D_{\rm kpc}$ the
distance in kpc, $\nu_{\rm GHz}$ the frequency in GHz, and S$_{\rm
\nu}$ the measured total flux density in Jy. Assuming a typical
electron temperature of T = 10$^4$ K, a distance of 4~kpc, and a  total
flux density of 2.7~Jy  as measured from the MAGPIS\footnote {http://third.ucllnl.org/gps/} image at 1.4~GHz, the total amount of
ionizing photons needed to keep this source ionized turns out to be
about N$_{uv} = (3.5\pm1.2) \times 10^{48} \rm ph~s^{-1}$. Assuming
errors of about 10\% in both the distance and the radio continuum flux
density, we conclude that the exciting star should be an O7V
\citep{mar05}.

\begin{figure}
\includegraphics[width=7.4cm]{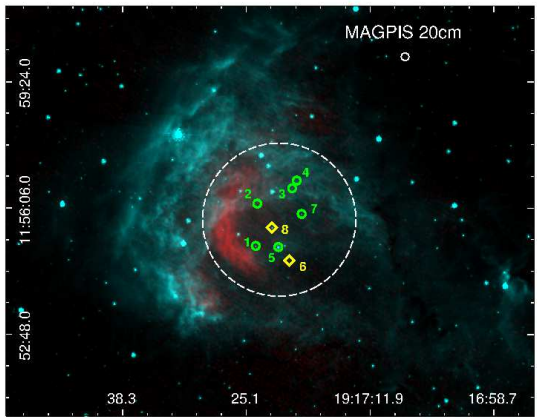}
\caption{Two-color composite image (8 $\mu$m = cyan and 20~cm =
red). The symbols represent the only eight point sources with
{\it JHK} near-infrared and RB optical bands measurements found in the region (dashed circle). The diamonds (sources \#6 and \#8)
indicate the location of the most likely candidates to be the exciting stars of G46. The beam of the 20 cm emission
from MAGPIS is included in the top right corner.}
\label{excitingstar}
\end{figure}

In a rough attempt to identify the star(s) candidate(s) responsible for
the ionized gas in the region we performed an optical and infrared
photometric study of the point sources in the area (see
Fig.\,\ref{excitingstar}) based on the astrometric UCAC3 Catalog
\citep{zac10}. Only sources with detection in the optical B and R
bands, and in the three near-infrared (NIR) {\it J,H,K} bands
extracted from the Two Micron All Sky Survey (2MASS)\footnote{2MASS is
a joint project of the University of Massachusetts and the Infrared
Processing and Analysis Center/California Institute of Technology,
funded by the National Aeronautics and Space Administration and the
National Science Foundation.}, were considered.  We found eight
sources inside a circle of size 4\m~ centered at the central coordinate of the
\hii~region.  
Their locations are shown in Fig.\,\ref{excitingstar}.
From the optical and infrared magnitudes we constructed their spectral
energy distribution (SED). We fit the available magnitudes (optical B
and R bands, the three {\it JHK} 2MASS bands, and the four {\it
Spitzer}-IRAC bands) using the Kurucz photospheric models
\citep{kur79} included in the tool developed by
\citet{rob07}\footnote{http://caravan.astro.wisc.edu/protostars/} to
obtain the effective temperature, T$_{eff}$, of each source. The
fitting tool requires the assumption of the visual extinction, A$_v$,
and the distance. We adopt a distance between 3.5 and
4.5~kpc. Regarding the A$_v$, we derived it for each source from their
{\it (J$-$H)} and {\it (H$-$K)} colors.  We assumed the interstellar
reddening law of \citet{rie85} (A$_J$ /A$_V$ = 0.282, A$_H$ /A$_V$
=0.175, and A$_K$ /A$_V$ = 0.112) and the intrinsic colors {\it
(J$-$H)$_0$} and {\it (H$-$K)$_0$} obtained by \citet{mart06}.

\begin{figure}
\includegraphics[width=9cm]{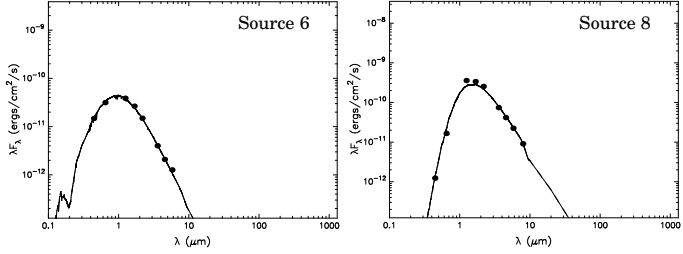}
\caption{Fitting of the spectral energy distribution for sources \#6 and \#8. The
black curves correspond to the best photospheric model of Kurucz (1979). The dots represent the data used for the fitting.}
\label{SED-Kurucz}
\end{figure}

Among the eight stars found in the region, only three are compatible
with a massive star located at the distance of G46.  The effective
temperatures of the massive candidates stars obtained from the fitting
are shown in Table \ref{tablestar}. Stars \#6 and \#8 have a T$_{eff}$
of about 30000~K and 35000~K, respectively, which agree with the
temperatures of O9.5V and O7V stars, respectively \citep{sch97}.
The models predict an effective temperature of $\sim$ 26000~K for
source \#3, suggesting that this star would be of a spectral type
later than B0. The fitted SEDs for sources \#6 and \#8 are shown in
Fig.\,\ref{SED-Kurucz}. For the remaining sources \#1, \#2, \#4, \#5
and \#7, the Kurucz's models fail to fit with confidence their optical
and NIR magnitudes as a massive star at about 4~kpc. These sources are
probably less massive foreground stars not related with G46. Thus,
stars \#6 and \#8 are the most probable candidates to be the exciting stars of the \hii~
region. Finally, the location of the star \#8 with respect to the 
radio continuum emission (red in Fig.\,\ref{excitingstar}), which looks like an incomplete ring,
seems to better explain the curved morphology observed in the
ionized gas distribution.

\subsection{Molecular gas analysis}
\label{secmolec}

The \hii~ region G46 is located in projection onto a border of the molecular cloud GRSMC G046.34-00.21 
catalogued by \citet{rath09} with an associated v$_{\rm LSR}$ of $54.3\pm3.9$ \ks. \citet{roman09} established
a kinematical distance of about 4 kpc for this molecular cloud, the same as for the \hii~ region G46, supporting 
that both objects are linked.  
Figure\,\ref{13cogrs} (left) displays the \3 J=1--0 integrated emission distribution in the  40 -- 62 \ks~velocity 
range. It can be seen that the \hii~region seems to be evolving on the northeastern edge of the cloud. 
The molecular gas appears located in 
projection onto the diffuse emission at 8 $\mu$m observed towards the southwest of the open border of
G46. 
It is noteworthy that there is no evidence of related molecular emission towards the east-northeast edge 
of the \hii~region, where its associated PAHs emission is thicker and more intense.  

The numerous studies of \hii~regions and their surroundings
usually show that the associated PAHs emission is more intense towards the borders of the \hii~regions with presence of molecular 
material. 
In other words, it is expected that an ionized bubble opens in the direction away from the cloud edge
because the ionized gas would be less confined and could stream-out into the lower
density ISM, forming what is known as a ``blister-type'' \hii~ region \citep{israel,tenorio}.
In the G46 case, however, the ring opens to the southwest, facing the elongated molecular cloud where apparently 
the density is higher. 
A possible explanation can be found from the morphology of the G46 radio continuum emission, which suggests
a stalling effect of the ionized gas against the photodissociation region (PDR) towards the northeast (see Fig.\,\ref{excitingstar}). 
This fact suggests 
that although there is no \3 J=1--0 detected towards this region, the density of the photodissociated gas is enough
to confine the ionized gas. The atypical thickness of the PDR towards this edge could be due to 
the presence of relatively low density gas which is expected towards the edge of a molecular cloud whereby the far-UV 
photons would have greater penetration in the material. By the other hand,  
the non detection of radio continuum emission towards the open border of G46 suggests that the ionized gas escapes from G46 
and dilutes in the ISM. The presence of diffuse emission at 8 $\mu$m towards the southwest of G46 supports this scenario. 

\subsubsection{The pillar-like features}
\label{pillars}

As above mentioned, about 10\m~towards the southwest of the 
\hii~region ($\sim$12 pc at the assumed distance of 4 kpc) it can be seen some pillar-like features emitting at 8 $\mu$m, 
which are pointing to the ionizing source 
of G46 and seem to be embedded in the molecular cloud GRSMC G046.34-00.21. 
Even though it seems to be a large distance, several pillars were found as far 
from the ionizing source as those analyzed in this work
(e.g in the Vulpecula rift; \citealt{billot10}). What it is really remarkable in this case is 
that these pillars-like features, resembling heaps and corrugations, are well outside of the \hii~region,
in contrary as is usually found: bubble-like structures with pillars inside or over their boundaries. Taking into account 
that the pillars point to the G46 open border, it is suggested
that they were produced by UV photons escaped from the \hii~region. 

\begin{figure}
\includegraphics[width=8cm]{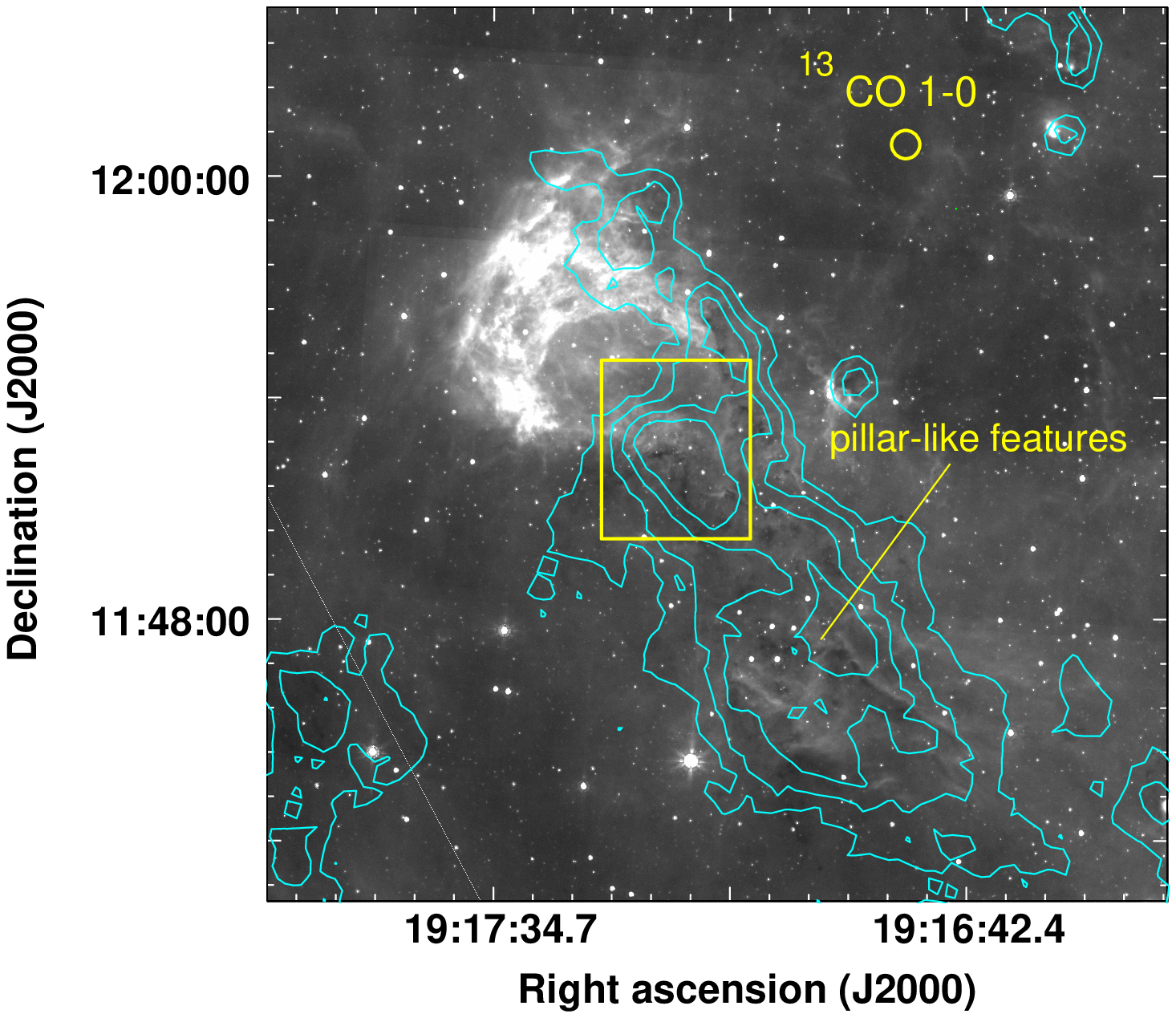}
\includegraphics[width=8cm]{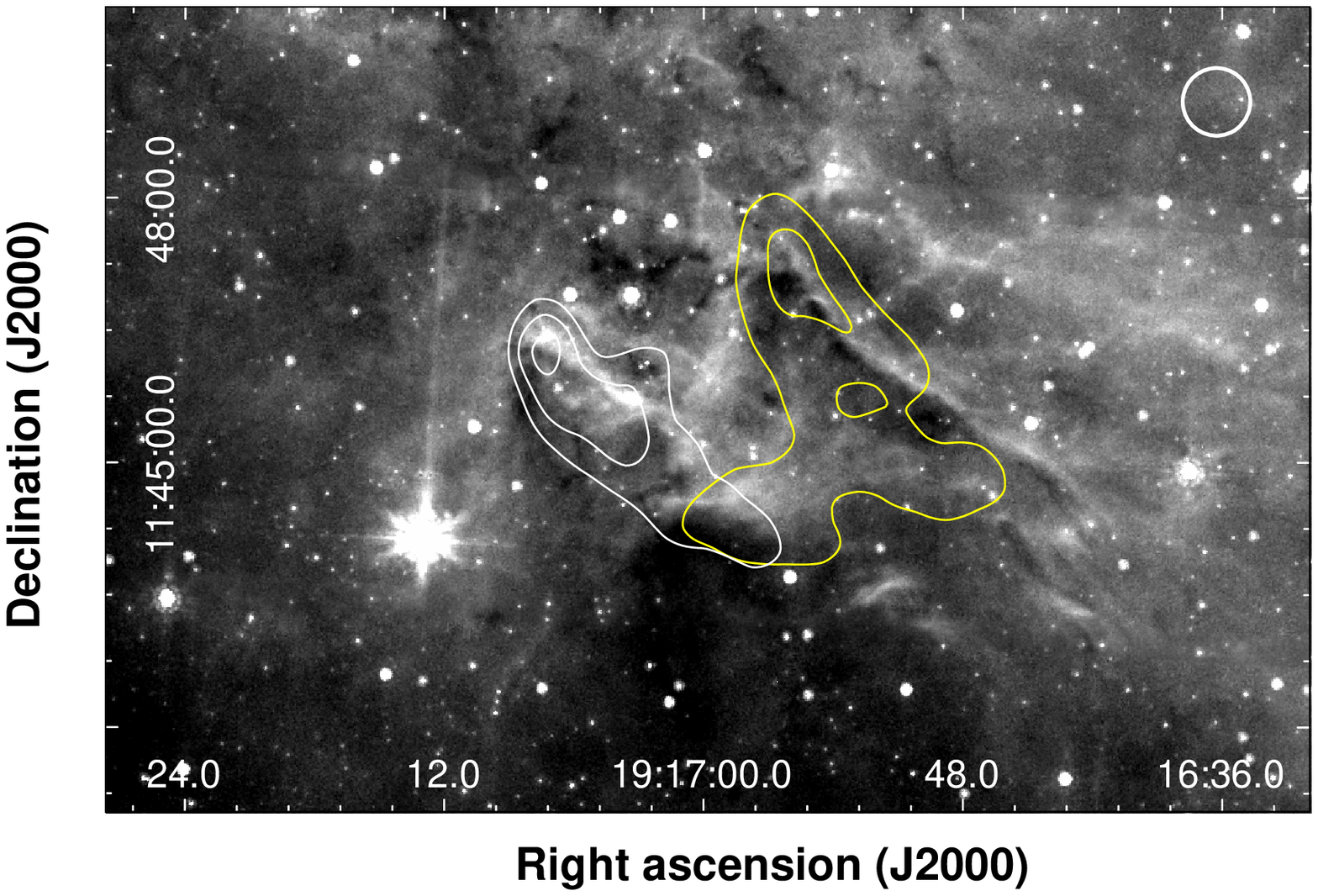}
\caption{Left: {\it Spitzer}-IRAC 8 $\mu$m emission with contours of the \3 J=1--0 emission integrated
between 40 and 62 \ks, with levels of 10, 15, 20, and 25 K \ks. The yellow circle is the beam of the observations.
The rms noise is about 1 K \ks. The rectangle represents the region studied with the JCMT data. Right: 
Integrated \3 J=1--0 from the GRS towards the pillar-like features observed in 8 $\mu$m. The white contours
represent the \3 emission integrated between 54.0 and 56.5 \ks~with levels of 4.5, 5.0, and 5.5 K \ks, while the yellow ones
represent the integration between 56.5 and 59.0 \ks~with levels of 4 and 5 K \ks. The beam of the molecular data is shown
in the top-right corner.}
\label{13cogrs}
\end{figure}

By inspecting the \3 J=1--0 towards the pillar-like features (Fig.\,\ref{13cogrs}-right), we find that there are
two molecular structures associated with them. The easternmost IR pillar-like feature has associated molecular gas 
between 54.0 and 56.5 \ks~(white contours in  Fig.\,\ref{13cogrs}-right), while the western one is related to a 
molecular structure ranging between 56.5 and 59.0 \ks~(yellow contours in Fig.\,\ref{13cogrs}-right).
The different velocity ranges for these molecular structures may either indicate that both features are located 
at slightly different positions along the line of sight or that they have a somewhat different kinematics.
It is remarkable the good agreement between the morphology of the molecular gas and the pillar-like features as 
seen in IR. Moreover, the molecular structures present a morphology consisting of a dense head with a less dense tail 
as is usually found and predicted by models and observations towards this kind of structures in the surroundings of \hii~ 
regions \citep{pound07,schuller06,pound05}. However, following the work of 
\citet{mackey10}, the appearance of these structures resembles  a previous evolutive stage in the formation
of pillars because it can not be appreciated a well-formed tail behind their heads.   
In what follow we characterize the interaction between the ionization radiation escaping from G46 and the pillar-like features.

In order to study the radiation influence over the tips of the pillar-like features and the possibilities of
triggered star formation via radiative driven implosion (RDI, e.g. \citealt{berto89}, \citealt{leflo97},
\citealt{kes03}) we evaluate the pressure
balance between the ionized gas stalling at the head of the pillars and the neutral gas of their interiors.
Assuming that the ionizing photons came from an O7V-type star located at the center of G46
(see Sect.\,\ref{star}), we use the predicted ionizing photon flux for an O7V star from \citet{sch97} and
the projected distance between the star and the pillars to roughly estimate the amount of UV photons arriving at the 
surface of the pillars in $\Phi_{pre} \sim 2 \times 10^{8}$ cm$^{-2}$~s$^{-1}$.
This value represents an upper limit due
to the fact that the projected distance between the star and the pillars is a lower limit to the actual distance between them.
Our predicted ionizing photon flux is similar to those measured towards several bright-rimmed clouds \citep{thompson04a,thompson04b}, which 
like the pillars, are molecular clouds sculpted by the radiation leaking from \hii~regions, and in many cases they have the same
large scale morphology as the pillars \citep{thompson04a}. 

Using the above obtained $\Phi_{pre}$ and following \citet{thompson04b} we estimate an upper limit for the 
electron density of about 64 cm$^{-3}$~for the expected ionized boundary
layer (IBL) at the tip of both pillar-like structures. This is almost 3 times greater than the critical value of 
$\sim25$ cm$^{-3}$ above which an IBL is able to develop around a cloud \citep{leflo94}. 
To obtain this value we consider that the pillar heads have a radius of 0.5\m~and assume an effective
thickness of the ionized boundary layer of $\eta = 0.1$. Then, using a typical sound speed of the ionized gas of 11.4 \ks, we obtain
that the pillars tips are supporting an external pressure of P$_{ext}/k \sim 2.1 \times 10^{6}$ cm$^{-3}$ K. It is important
to note that the obtained  P$_{ext}$~represents
strictly an upper limit because the predicted $\Phi$ is an upper limit and due to the used $\eta$ (see \citealt{thompson04b}).

On the other side, integrating the \3 J=1--0 emission over the area that contain each pillar head and using the typical
LTE formulae to derive the \3 and H$_{2}$ column densities by assuming a T$_{\rm ex} = 10$ K and [H$_{2}$/\3] $= 5 \times 10^{5}$
(see e.g. \citealt{yama99}), we estimate the mass of each pillar head in about 450 and 300 \msun~for the eastern and western one, respectively.
The area of each pillar head was assumed by taking into account the radius of their tips curvature, this is a circular area with
radius of 0.5\m~for both pillar heads. Thus, assuming spherical shapes, we estimate the molecular densities
in about $7.0 \times 10^{3}$ and $4.6 \times 10^{3}$ cm$^{-3}$, respectively. Then, considering the velocity interval in which each
pillar structure extends, we obtain the velocity dispersion $\sigma_{v} \sim 1.05$ \ks, which is in agreement with those predicted
by models of pillars formation \citep{grit10,dale12}. Finally, using the obtained densities and $\sigma_{v}$ we derive the internal
pressure for each pillar head in P$_{int}/k \sim 1.8 \times 10^{6}$, and $\sim 1.2 \times 10^{6}$ cm$^{-3}$ K for the eastern and western one,
respectively. As \citet{thompson04b} state, these pressure values are very likely underestimated because the \3 J=1--0 line underestimates
the true density and this molecule may be depleted by selective photodissociation at the boundary
of the clouds. Following these authors,
the internal pressure is likely underestimated by no more than a factor of 15, thus we conclude that  P$_{int}/k$
should be between $2 \times 10^{6}$ and $3 \times 10^{7}$ cm$^{-3}$ K. 

In conclusion, we obtain P$_{int}$ $>$ P$_{ext}$, which suggests that the diluted ionization front stalls at the pillar heads,
probably until the effects of mass evaporation and increasing recombination within the IBL raises the ionized
gas pressure to equilibrium with the interior pressure \citep{leflo94}. This result shows 
that it is unlikely that a shock is propagating farther into the molecular gas, discarding that the RDI mechanism is on going 
in the pillars interior.

\subsubsection{The molecular gas towards G46 open border}
\label{southclump}

\begin{figure}
\includegraphics[width=15cm]{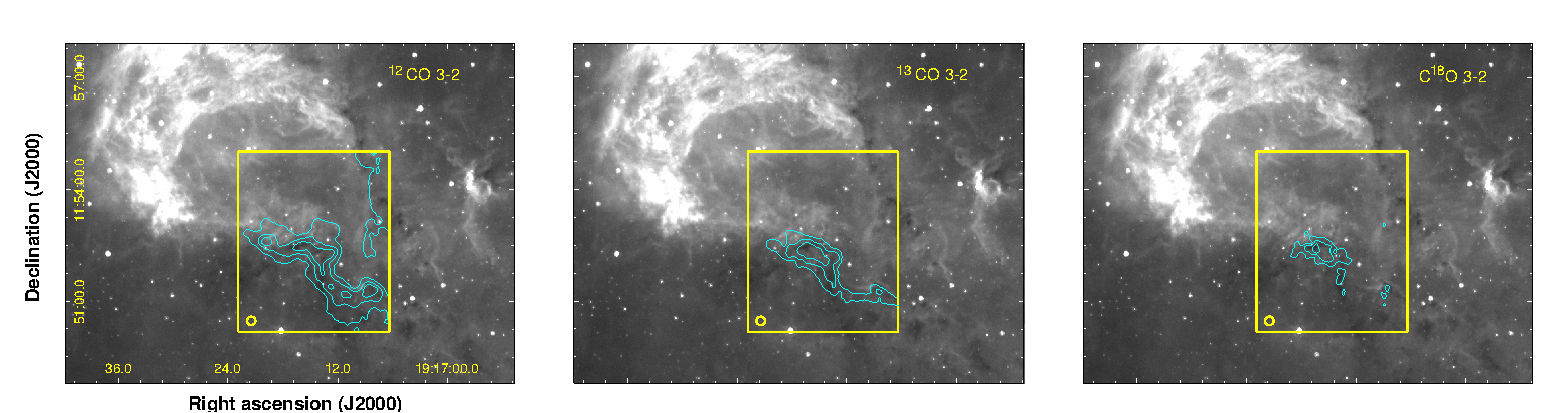}
\caption{\2, \3, and C$^{18}$O J=3--2 emission integrated between 40 and 60 \ks~displayed in contours 
over the {\it Spitzer}-IRAC 8 $\mu$m emission. The contours levels are 45, 58, and 72 K \ks; 13, 20, and 27 K \ks; and 3.3, 5.0, 
and 6.7 K \ks, for the \2, \3, and C$^{18}$O, respectively. The rms noises are about 2.5, 0.8, and 0.5 K \ks, respectively. 
The beam of the observations is shown at the bottom left corner of the surveyed region.}
\label{cojc}
\end{figure}

\begin{figure}
\includegraphics[width=8.5cm]{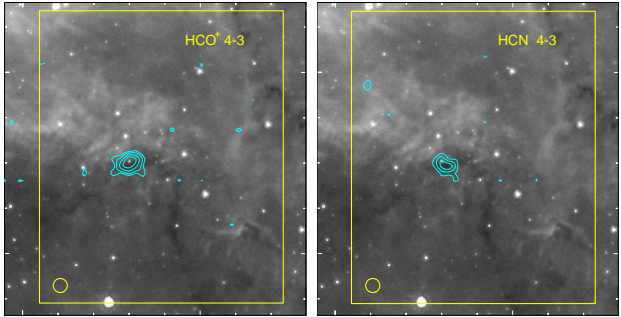}
\caption{{\it Spitzer}-IRAC 8 $\mu$m emission with contours of the J=4--3 transition of the \H~(left) and HCN (right) integrated
between 40 and 65 \ks. The contours levels are 2.3, 3.3, 5.0, and 8.3 K \ks~for the \H, and 1.2, 2.0, and 2.8 K \ks~for the HCN.
The rms noise is about 0.3 K \ks~for both emissions. The yellow rectangle represents the same area as shown in previous figures.}
\label{hcnco+}
\end{figure}

With the purpose of investigating the molecular material towards the open border of G46 
we used higher-angular resolution data of several molecular species extracted from the JCMT database acquired towards 
this region (see the rectangle in Fig.\,\ref{13cogrs}-left). 
Figure \ref{cojc} shows the integrated emission of the \2, \3, and C$^{18}$O J=3--2, while
Fig.\,\ref{hcnco+} displays the \H~and HCN J=4--3 integrated emission. 
The \2 and \3 J=3--2 emission show clumpy, elongated molecular features, with 
a high density filament extending towards the southwest of the mapped region. Towards 
the open border of G46 do not appear any considerable molecular emission, suggesting the presence of a pre-existing 
region with scarce molecular gas or that the UV-photons have carved the molecular cloud. In any case this can
be the path followed by the UV-photons escaped from G46 to reach the farther pillar-like structures.

The C$^{18}$O J=3--2, \H~and HCN J=4--3 emissions are concentrated in a small compact clump 
related to the infrared dark cloud IRDC 046.424-0.237 \citep{peretto09} and to the millimeter continuum source BGPS G046.427-00.237
(see the Bolocam Galactic Plane Survey v2; \citealt{ginsburg13}), suggesting that the emission lines emanate from a deeply 
embedded clump. Moreover, this structure is associated with the cold high-mass clump G46.43-0.24 \citep{wienen12} in
which the (1,1) and (2,2) ammonia lines were detected at v$_{\rm LSR} \sim$ 52.3 and 52.9 \ks, respectively.
The detection of these molecular species, mainly the HCN J=4--3 line that its critical density
can be between some $10^{6}$ and $10^{8}$ cm$^{-3}$ \citep{taka07,greve09}, indicates the presence of very high density gas. 
Taking into account that this clump is a potential site of star formation, in what follows we analyze it.

\begin{figure}
\includegraphics[width=0.8\textwidth]{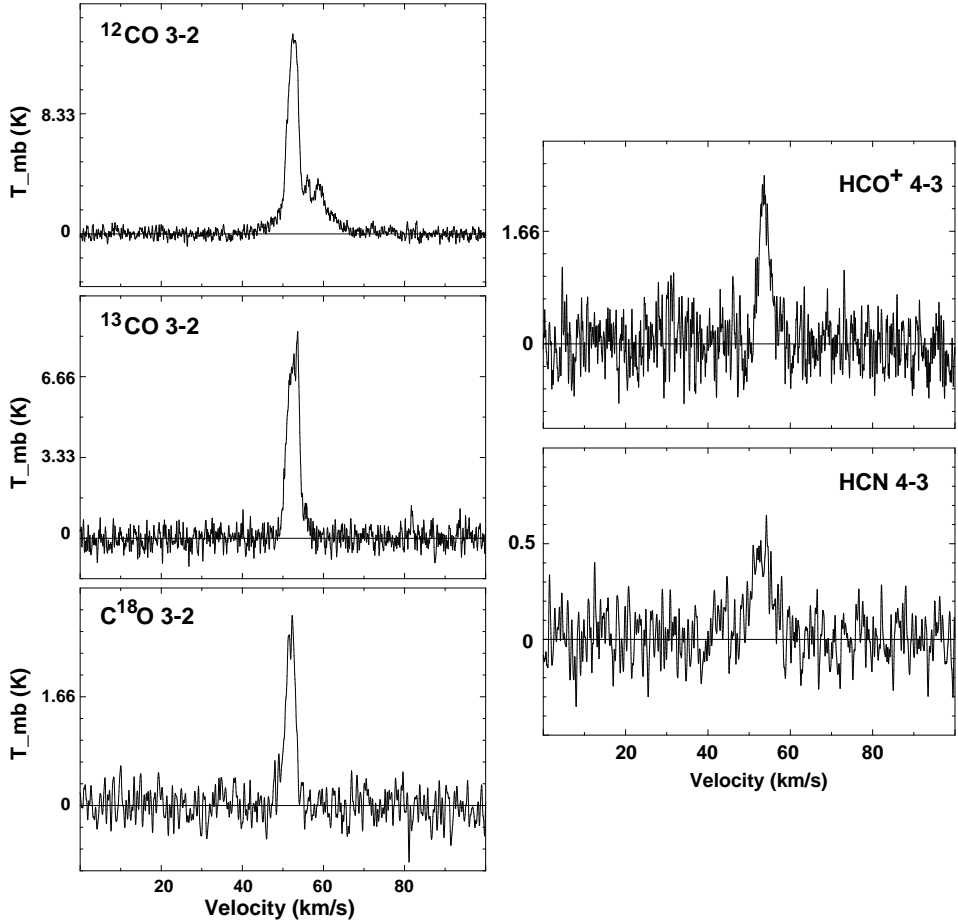}
\caption{Left: CO isotopes spectra. The rms noises of each spectrum are: 250, 400, and 200 mK, respectively. Right: \H~and HCN spectra.
The rms noises are 33 and 116 mK, respectively. All spectra were obtained towards the center of the dense molecular clump related to
the IRDC 046.424-0.237 and to the millimeter continuum source BGPS G046.427-00.237. }
\label{spectra}
\end{figure}

The line parameters of the observed molecular transitions towards the center of the clump are given in Table\,\ref{lines} 
as derived from Gaussian fits from the spectra shown in Fig.\,\ref{spectra}.
The \2 J=3--2 was fitted using three Gaussians, while the parameters of the others lines were obtained 
from single-components Gaussian fits which coincide in velocity with the main \2 component. 
The different velocity components observed in the \2 spectrum are
present almost in the whole region. Thus, we conclude that they correspond 
to different molecular components seen along the line of sight, reflecting the clumpiness in the region. 

In order to have a rough estimate of the molecular clump mass we assume local 
thermodynamic equilibrium (LTE). We
calculate the excitation temperature from
\begin{equation}
T_{ex}(3 \rightarrow  2) = \frac{16.59 {\rm K}}{{\rm ln}[1 + 16.59 {\rm K} / (T_{\rm max}(^{12}{\rm CO}) + 0.036 {\rm K})]}
\label{eq1}
\end{equation}
where $T_{\rm max}(^{12}{\rm CO}$) is the $^{12}$CO peak temperature towards the clump center at $\sim 52$ \ks, obtaining T$_{ex} \sim$ 18 K.
We derive the \3 and C$^{18}$O optical depths $\tau_{13}$ and $\tau_{18}$ from (e.g. \citealt{buckle10}):
\begin{equation}
\frac{^{13}{\rm T}_{mb}}{^{18}{\rm T}_{mb}} = \frac{1-exp(-\tau_{13})}{1-exp(-\tau_{13}/X)},
\label{eq2}
\end{equation}
where $^{13}$T$_{mb}$ and $^{18}$T$_{mb}$ are the peak temperatures of the \3 and C$^{18}$O 
J=3--2 line at the center of the region, and $X = 8.4$ is the assumed isotope
abundance ratio [\3/C$^{18}$O] \citep{frerking82,wilson99}, obtaining $\tau_{13} \sim 3.5$ and $\tau_{18} \sim 0.4$, which
indicate that the C$^{18}$O J=3--2 line appears moderately optically thin. Thus, we estimate its column density from:
\begin{equation}
{\rm N(C^{18}O)} = 8.26 \times 10^{13}~e^{\frac{15.81}{T_{ex}}}\frac{T_{ex}+0.88}{1-e^{\frac{-15.81}{T_{ex}}}} \frac{1}{J(T_{ex})-J(T_{\rm BG})} \int{{\rm T_{mb} ~dv}} 
\label{eq3}
\end{equation}
with
\begin{equation}
J(T) = \frac{h\nu/k}{exp(\frac{h\nu}{kT}) - 1}.
\label{eq4}
\end{equation}
To obtain the molecular hydrogen column density N(H$_{2}$) we assume an abundance ratio of
[H$_{2}$/C$^{18}$O] $= 5.88 \times 10^{6}$ \citep{frerking82,wilson99}. Finally the mass was derived from:
\begin{equation}
{\rm M} = \mu~m_{{\rm H}} \sum_{i}{\left[ D^{2}~\Omega_{i}~{\rm N_{\it i}(H_{2}}) \right]}, 
\label{eq5}
\end{equation}
where $\Omega$ is the solid angle subtended by the beam size, $m_{\rm H}$ is the hydrogen mass,
$\mu$, the mean molecular weight, is assumed to be 2.8 by taking into account a relative helium abundance
of 25 \%, and $D$ is the distance. Summation was performed over all beam positions on the molecular
structure observed in C$^{18}$O displayed in contours in Fig.\,\ref{cojc}. The obtained mass is about 375 \msun. 
Assuming an ellipsoidal volume with semi-axis of 45\s, 20\s, and 20\s~we derive a density 
of n $\sim 1.5 \times 10^{4}$ cm$^{-3}$. The density should be higher in the innermost region of the clump where 
the HCN J=4--3 line emanates, showing a density gradient. Using the deconvolved radius of about 0.5 pc calculated from:
\begin{equation}
{\rm R_{clump}} = \sqrt{\frac{S - {\rm beam~area}}{\pi}}
\label{radius}
\end{equation}
where $S$ is the area inside the clump, we appreciate that the above obtained mass value is slightly over the 
mass--size threshold for massive star formation in IRDCs presented in \citet{kauffmann10}, suggesting that massive YSOs can 
be formed within this clump.

Using the derived \H~and HCN J=4--3 parameters listed in Table\,\ref{lines} we
performed a non-LTE study of these molecular species with the code RADEX, which uses the mean escape probability approximation 
for the radiative transfer equation \citep{vander06}. 
Using the measured $\Delta$v we ran the code to fit T$_{\rm mb}$ and estimate the column densities. 
Taking into account that \citet{wienen12} measured from the ammonia lines a kinetic temperature
of T$_{\rm k} \sim 16$ K towards this clump, we fix this parameter in 20 K and assume densities between 10$^{5}$ 
and 10$^{7}$ cm$^{-3}$ to obtain the values presented in Table\,\ref{tradex}. From the obtained column densities 
we observe abundance HCN/\H~ratios of about 4.4, 3.8, and 1.2 for n$_{\rm H_{2}} = 10^{5}$, 10$^{6}$, and 10$^{7}$ cm$^{-3}$, respectively. 
In all cases we obtain a N(HCN)/N(\H) ratio larger than unity as found towards several clumps in the 
IRDC G48.66-0.22 \citep{pittan13} and the active star-forming region W49A \citep{roberts11}. 
As these authors point out, the steady-state chemical models for molecular species in gas-phase predict HCN $>$ \H~only
for T$_{\rm k} <$ 25 K, with a density of n$_{\rm H_{2}} = 10^{6}$ cm$^{-3}$, which is consistent with our results.

\subsection{Star formation around G46}

Finally, given that the scenario  is very favorable to initiate the formation of new generations of stars, 
we analyze in this section the existence and properties of young stellar objects in all the investigated area.

\subsubsection{Identification of YSOs}
\label{secYSOs}

Young stellar objects (YSOs) always show an excess in the infrared emission. The level of excess in the infrared can be effectively 
used for discriminating YSOs from field stars and distinguishing different evolutionary stages. At a considerably early 
evolutionary stage, protostars are mostly embedded in dust envelopes, they exhibit large excess of infrared emission and 
an infrared spectral index $\alpha_{\rm{IR}}>-0.3$ indicative of flat or ascending spectral energy distribution (SED) 
at wavelength longward of 2 $\mu$m \citep{lad87,gre94}. For pre-main sequence (PMS) stars which possess optically thick disks, 
the SEDs tend to descend and the infrared spectral indices are in the range $-1.6<\alpha_{\rm{IR}}<-0.3$. Finally  ``transition disk'' 
sources are more evolved YSOs, where the inner part of the disks have been cleared by photoevaporation of the central stars or by 
planet forming processes. For such YSOs it is expected to detect IR excess at  
wavelengths longer than 16 $\mu$m \citep{str89}.
These properties of YSOs make photometric observations in the near- to mid-infrared plausible to discriminate them from field 
stars. Different color-based source identification and classification schemes have been developed and verified in the practice. 
In this section, we identify potential YSOs following the scheme proposed by \citet{gut09}. The resulting YSOs are classified into Class I 
(protostars, including Class 0, Class I, and ``flat spectrum'' sources), Class II, and ``transition disk'' sources.

There are several kinds of contaminants that could be misidentified as YSOs in our original sample. Extragalactic 
contaminations could stem from star-forming galaxies and broad-line active galactic nuclei (AGN), which show PAH feature 
emission yielding very red 5.8 and 8.0 $\mu$m colors \citep{ste05,gut08}. In our own Galaxy, unresolved knots of shock emission      
and resolved PAH emission are often detected in the IRAC bands, yielding additional contaminations. Following the techniques presented in 
\citet{gut09} we exclude contaminants, and then identify and categorize potential YSOs as described as follows.

In the first phase, only sources with valid detections in all four IRAC bands have been considered. Any source fulfilling 
color criteria of $[3.6]-[4.5]>0.7$ and $[4.5]-[5.8]>0.7$ is regarded as a Class I YSO.  In the remaining pool, 
Class II sources have been picked out based on the constraints of {\it i}) $[3.6]-[4.5]-\sigma_1>0.15$, {\it ii}) 
$[3.6]-[5.8]-\sigma_2>0.35$, {\it iii}) $[4.5]-[8.0]-\sigma_3>0.5$, and {\it iv}) 
$[3.6]-[5.8]+\sigma_2\le\frac{0.14}{0.04}\times(([4.5]-[8.0]-\sigma_3)-0.5)+0.5$. 
Here, $\sigma_1=\sigma([3.6]-[4.5])$, $\sigma_2=\sigma([3.6]-[5.8])$,  and $\sigma_3=\sigma([4.5]-[8.0])$ are combined errors, 
added in quadrature.

In the second phase, sources with 24 \micron~data in the remaining pool are reexamined. Sources with colors of $[5.8]-[24]>2.5$ 
or $[4.5]-[24]>2.5$ are classified to be ``transition disks''. With the potential contaminants and YSOs identified 
above excluded, there still remain some sources with bright 24 \micron~emission. For sources lacking valid  photometric data 
in one or more of the IRAC bands, they are picked out as additional Class I type YSOs once fulfilling $[24]>7$ and $[X]-[24]>4.5$ mag, 
where $[X]$ is the longest wavelength IRAC detection that we have.

The classification scheme of \citet{gut09} includes three phases, using 2MASS
data to identify more  YSOs in addition to the aforementioned methods. However, in our case as G46 is 
much farther than the sources in \citet{gut09}, 2MASS 
photometry would be severely affected by foreground interstellar extinction and the use of these near-IR data to identify YSOs would induce 
heavy contamination from foreground field stars. Thus, we do not use the near-IR bands in our YSOs search.

\begin{figure}
\includegraphics[width=0.8\textwidth]{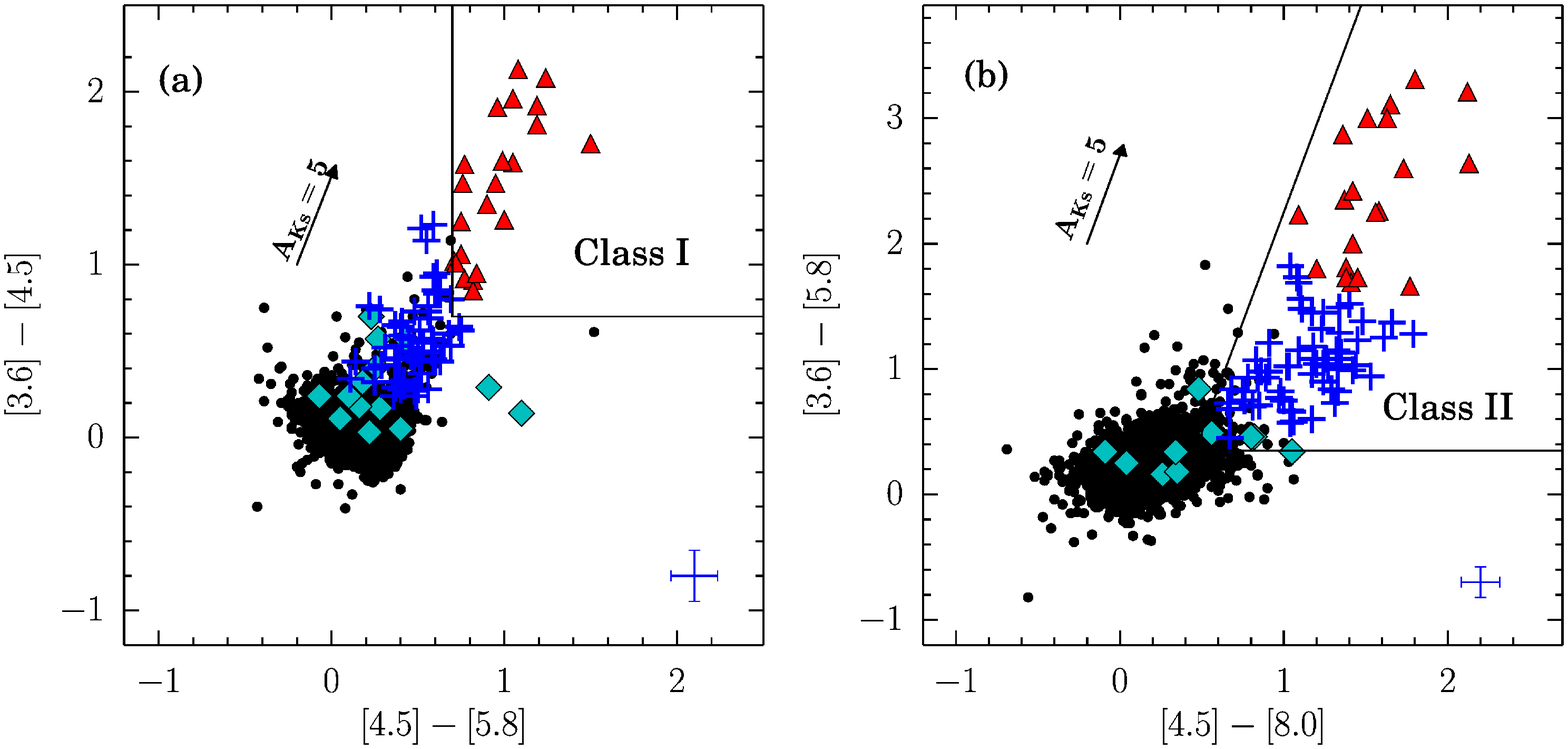}
\caption{IRAC color-color diagrams. Class I, Class II, and ``transition disk'' YSOs are marked using red triangles,
blue plus symbols, and cyan diamonds, respectively. Mean errors of colors are presented in the bottom right
corner of each panel. The definition of loci are adopted from \citet{gut09}.}
\label{fig-cc-irac}
\end{figure}

\begin{figure}
\includegraphics[width=0.45\textwidth]{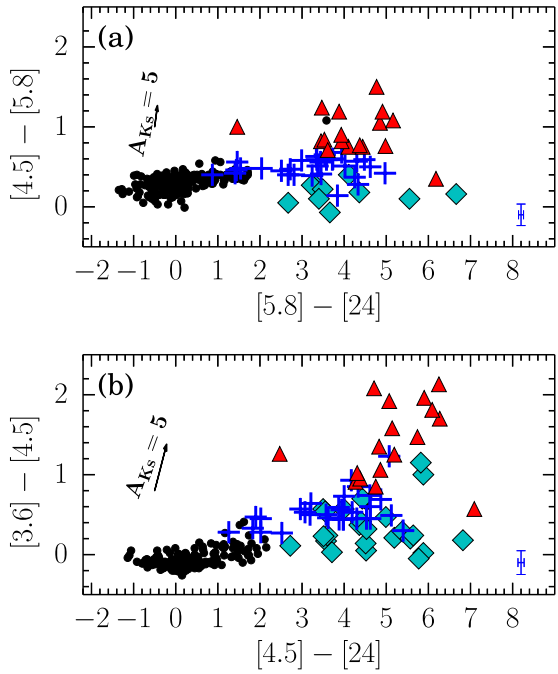}
\caption{IRAC-MIPS color-color diagrams. Class I, Class II, and ``transition disk'' YSOs are marked using red triangles,
blue plus symbols, and cyan diamonds, respectively.  Mean errors of colors are presented in the bottom right corner of each panel.}
\label{fig-cc-mips}
\end{figure}

The above identification procedures have resulted in 106 YSOs. And they are classified into 22 Class I, 60 Class II, 
and 24 ``transition disks'' objects.
A summary of the results is given in Table \ref{tb-count}. We note that the color based classification scheme would lead to 
misidentification. A Class II YSO viewed at high inclination would show features resembling that of a Class I source. 
An edge-on Class I YSO can have similar infrared color of a Class 0 source. Thus, all Class I, II, and ``transition disk'' 
sources identified in this paper are YSO candidates. The distribution of these YSO candidates on distinct color-color spaces 
are presented in Figs. \,\ref{fig-cc-irac} and \,\ref{fig-cc-mips}.
It is important to note that we also use the point sources extracted from \citet{gut15} to identify YSO candidates and reached identical results.

\subsubsection{Active Star Formation in GRSMC G046.34-00.21}

\begin{figure}
\includegraphics[width=0.45\textwidth]{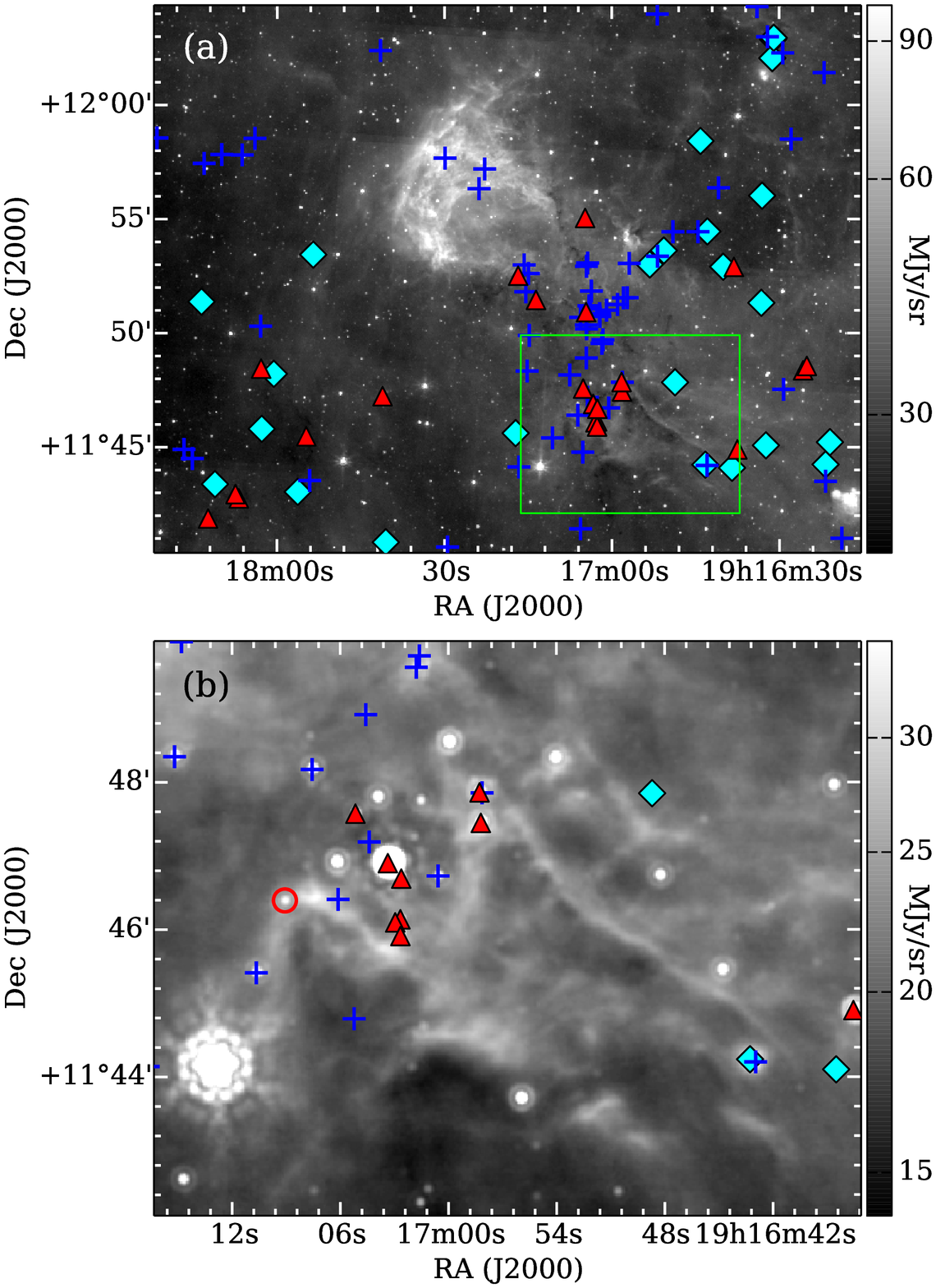}
\caption{(a) Distribution of YSOs. Class I, Class II, and ``Transition Disk'' YSOs are marked using red triangles,
blue plus symbols, and cyan diamonds, respectively. The grayscale shows emission at 8.0 \micron. (b) A closeup
view of the pillar-like features region. The background presents emission at 24 \micron.}
\label{fig:yso-dis}
\end{figure}

Figure \ref{fig:yso-dis}(a) shows the large scale spatial distribution of the 106 YSOs identified in the region. 
Among the 82 Class I and Class II YSOs more than 40 (about 50 percent) are located in projection onto the molecular cloud 
GRSMC G046.34-00.21 (see Fig. \ref{13cogrs}) which fills only the 20 percent of the whole survey field. 
Such a concentration of YSOs is suggestive of active star formation taking place in the cloud. By the other hand, 
from the 24 ``transition disk'' found in the whole field, no one is detected in association with the molecular gas, indicating 
the relatively young nature of the YSOs embedded in this cloud.
Interestingly all YSOs associated with the molecular cloud are placed in projection between the open border of G46
and the head of the pillars. Moreover, there are not young sources in the farthest part of the cloud 
behind the heads of these pillars (see Fig. \ref{fig:yso-dis}(b)). 

From Figure \ref{fig:yso-dis}(a) it can be appreciated only two concentrations of young objects over the whole field,
one centered at 19\hh16\mm30\ss, $+$11\deg51\m00\s~(J2000), consisting of 15 Class II type sources and
the other one centered at 19\hh17\mm00\ss, $+$11\deg47\m00\s~(J2000) mostly composed by Class I type
YSOs (9 sources), which are located just ahead the pillars-like features. 
A closeup view of the pillar-like features region is shown in Figure \ref{fig:yso-dis}(b). There is a point
source revealed at 24 $\mu$m which is too faint at IRAC bands and not listed in the GLIMPSE
catalog. We have marked this 24 $\mu$m point source using a red circle in Figure \ref{fig:yso-dis}(b). Weak
at short wavelengths, this point source would be younger than other ones and could be a Class 0 candidate.

The fact that the Class II
concentration is located closer to the open border of G46 than the Class I group strongly
suggests an age gradient in the YSOs distribution. By the other hand, the absence of young sources inside 
and behind the pillar-like structures could show that the \hii~region influence has not reached the molecular 
material behind the pillars, which is in agreement with the result of Section \ref{pillars}. 
We suggest a scenario where the propagation of ionization radiation escaping from the \hii~region has triggered 
the star formation observed in the molecular cloud through RDI mechanism and then has stalled at the 
surface of the pillars heads.  
We can not discard that a growing density of the IBL reaches equilibrium with that of the cloud and thus, the shock front 
will continue their propagation into the cloud \citep{thompson04a}.   

Additionally, in the region surveyed with the JCMT (see yellow rectangle in Fig.\,\ref{13cogrs}) there are 5 YSO candidates. One of them,
a Class I source, lies exactly at the center of the dense molecular clump mapped with the \H~and HCN J=4--3 (see Fig.\,\ref{hcnco+}),
which is in agreement with the molecular gas conditions studied in Sec.\,\ref{southclump}.

\section{Conclusions}

We have analyzed the  \hii~region G46.5-0.2, its molecular environment and the young stellar objects placed in a wide 
region around it, searching for evidence of their physical connection and the possibility of induced star formation. 
We found a rich combination of mutual influences, underscoring the important role of \hii~regions in favoring the formation 
of new stars not only on their immediate vicinity but also farther than a distance equivalent to its own radius.

We characterized that the \hii~region, located at about 4 kpc and with a radius of $\sim 4.6$ pc, is probably excited by an O7V star. 
Its shape resembles a horse-shoe, with the bright, thick border towards the east-northeast and the open portion towards the southwest side.  
When mapping the distribution of the molecular gas associated with G46 we found that the \hii~region is located close to the edge of the 
GRSMC G046.34-00.21 molecular cloud and, curiously, G46 instead of opening in the direction of lower ambient density it does exactly in the 
opposite direction, towards  the cloud. 
Filamentary structure in the molecular cloud is observed, 
particularly in the $^{12}$CO J=3--2 and $^{13}$CO J=3--2 maps. 
Towards the open border of G46 do not appear any considerable molecular emission, 
suggesting the presence of a pre-existing region with scarce molecular gas or 
that the UV-photons have carved the molecular cloud.
Besides, close to the end of the observed filamentary structures in the cloud, the infrared images reveal the existence of pillar-like 
structures pointing towards the \hii~region open border, which are associated with some narrow molecular filaments characterized 
using the $^{13}$CO J=1--0 line. From a pressure balance study we found that the internal pressure  
of the neutral gas in the pillar-like features heads are larger than the external pressure due to the ionized gas stalling at their tips,  
discarding that the RDI mechanism is on going.
In addition, we analyzed a compact molecular clump  
located over the south border of G46, obtaining a mass of $\sim$ 375 \msun~ and a density of about $10^4$ cm$^{-3}$, with a higher density
towards its center, where the HCN J=4--3 line is detected.

Taking into account that the expansion of the \hii~region itself, and the injection of extra energy into the molecular cloud can drive 
turbulence and trigger star formation we searched for candidates of YSOs and classified them according to their evolutionary stage. 
We identified two main concentrations of young objects over the whole region,
one, closer to the G46 open border consisting of Class II type sources, and other one mostly composed by Class I type YSOs located 
just ahead the pillars-like features strongly suggesting  an age gradient in the YSOs distribution.

\acknowledgements
We wish to thank the anonymous referee for her/his useful comments and corrections. 
S.P., M.O., G.D., A.P., and E.G. are members of the {\sl Carrera del 
Investigador Cient\'\i fico} of CONICET, Argentina. G.D. acknowledges hospitality at 
the National Astronomical Observatories (Chinese Academy of Sciences) in Beijing.
This work was partially supported by Argentinian grants awarded by CONICET, ANPCYT and UBA (UBACyT).
This work is partially supported by the China Ministry of Science and Technology through 
grant of 2010DFA02710 (the China-Argentina Radio Telescope project) and NSFC through grants 
of 110732027, 11373009, 11433008, and 11403040. J.-H. Y and Y. F. H are supported by the Young Research Grant of NAOC.

\clearpage

\clearpage

\begin{table}
\begin{center}
\tiny
\caption{Exciting star(s) candidate(s) for G46.}
\label{tablestar}
\begin{tabular}{ccccccccccc}
\tableline\tableline
\# & 2MASS Designation & $J$ & $H$ & $K$ & B & R &  A$_v$[mag] & T$_{eff}$[K] & $\chi^2_{\rm per point}$&
Spectral type\\ \tableline
3&19172025+1156369&12.508&12.130&11.942&16.80&14.70&4-5&$\sim$26000&1.08&later than B0\\
6&19172053+1154438&12.498&12.108&11.9815&15.70&14.10&3.5-4.5&$\sim$30000&1.28&$\sim$O9.5V\\
8&19172239+1155353&10.094&9.353& 8.914&18.40&14.80&4-5&$\sim$35000&1.16&$\sim$[O7-O7.5]V\\
\tableline
\end{tabular}
\end{center}
\end{table}

\begin{table}
\begin{center}
\caption{Line parameters for the molecular lines towards the center of the southern molecular clump.}
\label{lines}
\begin{tabular}{lccc}
\tableline
\tableline
Molecular line            &  T$_{\rm mb}$ peak & v$_{\rm LSR}$    & $\Delta$v (FWHM)  \\
                          &   (K)             &   (\ks)         &  (\ks)            \\
\tableline
\noalign{\smallskip}
\2 (3--2)                 &   13.90$\pm$0.80     &  52.50$\pm$0.05  &  2.95$\pm$0.15         \\
                          &   1.93$\pm$0.33     &  56.10$\pm$0.15  &  0.95$\pm$0.25         \\
                          &   2.92$\pm$0.25     &  58.70$\pm$0.30  &  6.10$\pm$0.80          \\
\3 (3--2)                 &   7.83$\pm$0.25     &  52.60$\pm$0.01  &  3.47$\pm$0.15         \\
C$^{18}$O (3--2)          &   2.75$\pm$0.33     &  51.80$\pm$0.15  &  2.80$\pm$0.40         \\
HCO$^{+}$ (4--3)          &   2.16$\pm$0.15     &  53.70$\pm$0.04  &  3.15$\pm$0.10        \\
HCN (4--3)                &   0.45$\pm$0.12     &  53.10$\pm$0.40  &  5.50$\pm$1.00        \\
\tableline
\end{tabular}
\end{center}
\end{table}

\begin{table}
\begin{center}
\caption{Radex results from the \H~and HCN J=4--3 lines using T$_{\rm k} = 20$ K and the indicated n$_{\rm H_{2}}$.}
\label{tradex}
\begin{tabular}{cccccc}
\tableline
\tableline
input n$_{\rm H_{2}}$      & N(\H)    &  $\tau_{\rm HCO^{+}}$   & N(HCN)         & $\tau_{\rm HCN}$ &  $X_{\rm HCN/HCO^{+}}$    \\
(cm$^{-3}$)  &   (cm$^{-2}$)         &                          &   (cm$^{-2}$)  &                  &   \\
\tableline
\noalign{\smallskip}
10$^{5}$       &   1.9 $\times 10^{14}$  &  7.50 &   8.4 $\times 10^{14}$   & 4.50  &   4.42    \\
10$^{6}$       &   1.7 $\times 10^{13}$  &  0.70 &   6.5 $\times 10^{13}$   & 0.37  &   3.82   \\
10$^{7}$       &   5.4 $\times 10^{12}$  &  0.14 &   6.8 $\times 10^{12}$   & 0.09  &   1.25    \\
\tableline
\end{tabular}
\end{center}
\end{table}

\begin{table}
\begin{center}    
    \caption{Source Counts in the YSO Search}
    \label{tb-count}
    \begin{tabular}{lr}
    \tableline\tableline
    \multicolumn{1}{c}{Sources} &   \multicolumn{1}{c}{G46 Field} \\
    \tableline
    In GLIMPSE Catalog              & 19210   \\
    Valid in all Four IRAC Bands\tablenotemark{a} & 4491     \\
    With 24 \micron~data            & 338    \\
    Rejected Contaminants           & 883     \\
    Class I Type                    & 22       \\
    Class II Type                   & 60        \\
    Transition Disks                & 24       \\
    Total YSO Candidates            & 106 \\
    \tableline
    \end{tabular}
    \tablenotetext{a}{Sources with photometric uncertainties no larger than 0.2 mag in all four IRAC bands.}
\end{center}
\end{table}


\begin{thebibliography}{}

\bibitem[{{Anderson} \& {Bania}(2009)}]{anderson09}
{Anderson}, L.~D. \& {Bania}, T.~M. 2009, \apj, 690, 706


\bibitem[{{Benjamin} {et~al.}(2003){Benjamin}, {Churchwell}, {Babler}, {Bania},
  {Clemens}, {Cohen}, {Dickey}, {Indebetouw}, {Jackson}, {Kobulnicky},
  {Lazarian}, {Marston}, {Mathis}, {Meade}, {Seager}, {Stolovy}, {Watson},
  {Whitney}, {Wolff}, \& {Wolfire}}]{ben03}
{Benjamin}, R.~A., {Churchwell}, E., {Babler}, B.~L., {et~al.} 2003, \pasp,
  115, 953

\bibitem[{{Bertoldi}(1989)}]{berto89}
{Bertoldi}, F. 1989, \apj, 346, 735

\bibitem[Billot et al.(2010)]{billot10} Billot, N., 
Noriega-Crespo, A., Carey, S., et al.\ 2010, \apj, 712, 797 

\bibitem[Bisbas et al.(2011)]{bisbas11} Bisbas, T.~G., 
W{\"u}nsch, R., Whitworth, A.~P., Hubber, D.~A., 
\& Walch, S.\ 2011, \apj, 736, 142 

\bibitem[Buckle et al.(2009)]{buckle09} Buckle, J.~V., Hills, 
R.~E., Smith, H., et al.\ 2009, \mnras, 399, 1026 


\bibitem[{{Buckle} {et~al.}(2010){Buckle}, {Curtis}, {Roberts}, \&
  {et~al.}}]{buckle10}
{Buckle}, J.~V., {Curtis}, E.~I., {Roberts}, J.~F., \& {et~al.} 2010, \mnras,
  401, 204

\bibitem[{{Carey} {et~al.}(2009){Carey}, {Noriega-Crespo}, {Mizuno}, {Shenoy},
  {Paladini}, {Kraemer}, {Price}, {Flagey}, {Ryan}, {Ingalls}, {Kuchar},
  {Pinheiro Gon{\c c}alves}, {Indebetouw}, {Billot}, {Marleau}, {Padgett},
  {Rebull}, {Bressert}, {Ali}, {Molinari}, {Martin}, {Berriman}, {Boulanger},
  {Latter}, {Miville-Deschenes}, {Shipman}, \& {Testi}}]{car09}
{Carey}, S.~J., {Noriega-Crespo}, A., {Mizuno}, D.~R., {et~al.} 2009, \pasp,
  121, 76

\bibitem[{{Chaisson}(1976)}]{cha76}
{Chaisson}, E.~J. 1976, in Frontiers of Astrophysics, ed. {E.~H.~Avrett},
  259--351

\bibitem[Dale et al.(2012)]{dale12} Dale, J.~E., Ercolano, B., 
\& Bonnell, I.~A.\ 2012, \mnras, 424, 377 


\bibitem[{{Deharveng} {et~al.}(2008){Deharveng}, {Lefloch}, {Kurtz}, {Nadeau},
  {Pomar{\`e}s}, {Caplan}, \& {Zavagno}}]{deha08}
{Deharveng}, L., {Lefloch}, B., {Kurtz}, S., {et~al.} 2008, \aap, 482, 585

\bibitem[{{Dirienzo} {et~al.}(2012){Dirienzo}, {Indebetouw}, {Brogan},
  {Cyganowski}, {Churchwell}, \& {Friesen}}]{diri12}
{Dirienzo}, W.~J., {Indebetouw}, R., {Brogan}, C., {et~al.} 2012, \aj, 144, 173

\bibitem[{{Elmegreen} {et~al.}(1995){Elmegreen}, {Kimura}, \& {Tosa}}]{elme95}
{Elmegreen}, B.~G., {Kimura}, T., \& {Tosa}, M. 1995, \apj, 451, 675

\bibitem[{{Fazio} {et~al.}(2004){Fazio}, {Hora}, {Allen}, {Ashby}, {Barmby},
  {Deutsch}, {Huang}, {Kleiner}, {Marengo}, {Megeath}, {Melnick}, {Pahre},
  {Patten}, {Polizotti}, {Smith}, {Taylor}, {Wang}, {Willner}, {Hoffmann},
  {Pipher}, {Forrest}, {McMurty}, {McCreight}, {McKelvey}, {McMurray}, {Koch},
  {Moseley}, {Arendt}, {Mentzell}, {Marx}, {Losch}, {Mayman}, {Eichhorn},
  {Krebs}, {Jhabvala}, {Gezari}, {Fixsen}, {Flores}, {Shakoorzadeh}, {Jungo},
  {Hakun}, {Workman}, {Karpati}, {Kichak}, {Whitley}, {Mann}, {Tollestrup},
  {Eisenhardt}, {Stern}, {Gorjian}, {Bhattacharya}, {Carey}, {Nelson},
  {Glaccum}, {Lacy}, {Lowrance}, {Laine}, {Reach}, {Stauffer}, {Surace},
  {Wilson}, {Wright}, {Hoffman}, {Domingo}, \& {Cohen}}]{faz04}
{Fazio}, G.~G., {Hora}, J.~L., {Allen}, L.~E., {et~al.} 2004, \apjs, 154, 10

\bibitem[{{Frerking} {et~al.}(1982){Frerking}, {Langer}, \&
  {Wilson}}]{frerking82}
{Frerking}, M.~A., {Langer}, W.~D., \& {Wilson}, R.~W. 1982, \apj, 262, 590


\bibitem[{{Ginsburg} {et~al.}(2013){Ginsburg}, {Glenn}, {Rosolowsky},
  {Ellsworth-Bowers}, {Battersby}, {Dunham}, {Merello}, {Shirley}, {Bally},
  {Evans}, {Stringfellow}, \& {Aguirre}}]{ginsburg13}
{Ginsburg}, A., {Glenn}, J., {Rosolowsky}, E., {et~al.} 2013, \apjs, 208, 14

\bibitem[{{Greene} {et~al.}(1994){Greene}, {Wilking}, {Andre}, {Young}, \&
  {Lada}}]{gre94}
{Greene}, T.~P., {Wilking}, B.~A., {Andre}, P., {Young}, E.~T., \& {Lada},
  C.~J. 1994, \apj, 434, 614

\bibitem[Greve et al.(2009)]{greve09} Greve, T.~R., 
Papadopoulos, P.~P., Gao, Y., \& Radford, S.~J.~E.\ 2009, \apj, 692, 1432 



\bibitem[Gritschneder et al.(2010)]{grit10} Gritschneder, M., 
Burkert, A., Naab, T., \& Walch, S.\ 2010, \apj, 723, 971 

\bibitem[Gutermuth 
\& Heyer(2015)]{gut15} Gutermuth, R.~A., \& Heyer, M.\ 2015, \aj, 149, 64 



\bibitem[{{Gutermuth} {et~al.}(2009){Gutermuth}, {Megeath}, {Myers}, {Allen},
  {Pipher}, \& {Fazio}}]{gut09}
{Gutermuth}, R.~A., {Megeath}, S.~T., {Myers}, P.~C., {et~al.} 2009, \apjs,
  184, 18

\bibitem[{{Gutermuth} {et~al.}(2008){Gutermuth}, {Myers}, {Megeath}, {Allen},
  {Pipher}, {Muzerolle}, {Porras}, {Winston}, \& {Fazio}}]{gut08}
{Gutermuth}, R.~A., {Myers}, P.~C., {Megeath}, S.~T., {et~al.} 2008, \apj, 674,
  336

\bibitem[{{Israel}(1978)}]{israel}
{Israel}, F.~P. 1978, \aap, 70, 769

\bibitem[{{Jackson} {et~al.}(2006){Jackson}, {Rathborne}, {Shah}, {Simon},
  {Bania}, {Clemens}, {Chambers}, {Johnson}, {Dormody}, {Lavoie}, \&
  {Heyer}}]{jackson06}
{Jackson}, J.~M., {Rathborne}, J.~M., {Shah}, R.~Y., {et~al.} 2006, \apjs, 163,
  145

\bibitem[Kauffmann 
\& Pillai(2010)]{kauffmann10} Kauffmann, J., \& Pillai, T.\ 2010, \apjl, 723, L7 

\bibitem[{{Kessel-Deynet} \& {Burkert}(2003)}]{kes03}
{Kessel-Deynet}, O. \& {Burkert}, A. 2003, \mnras, 338, 545

\bibitem[{{Kuchar} \& {Bania}(1994)}]{kuchar94}
{Kuchar}, T.~A. \& {Bania}, T.~M. 1994, \apj, 436, 117

\bibitem[{{Kurucz}(1979)}]{kur79}
{Kurucz}, R.~L. 1979, \apjs, 40, 1

\bibitem[{{Lada}(1987)}]{lad87}
{Lada}, C.~J. 1987, in IAU Symposium, Vol. 115, Star Forming Regions, ed.
  M.~{Peimbert} \& J.~{Jugaku}, 1--17

\bibitem[Lefloch 
\& Lazareff(1994)]{leflo94} Lefloch, B., \& Lazareff, B.\ 1994, \aap, 289, 559 

\bibitem[Lefloch et 
al.(1997)]{leflo97} Lefloch, B., Lazareff, B., \& Castets, A.\ 1997, \aap, 324, 249 


\bibitem[{{Lockman}(1989)}]{lockman89}
{Lockman}, F.~J. 1989, \apjs, 71, 469

\bibitem[{{Martins} \& {Plez}(2006)}]{mart06}
{Martins}, F. \& {Plez}, B. 2006, \aap, 457, 637

\bibitem[{{Martins} {et~al.}(2005){Martins}, {Schaerer}, \& {Hillier}}]{mar05}
{Martins}, F., {Schaerer}, D., \& {Hillier}, D.~J. 2005, \aap, 436, 1049

\bibitem[Mackey 
\& Lim(2010)]{mackey10} Mackey, J., \& Lim, A.~J.\ 2010, \mnras, 403, 714

\bibitem[Ma{\'{\i}}z-Apell{\'a}niz et al.(2004)]{maiz04} 
Ma{\'{\i}}z-Apell{\'a}niz, J., Walborn, N.~R., Galu{\'e}, H.~{\'A}., 
\& Wei, L.~H.\ 2004, \apjs, 151, 103 

\bibitem[Ma{\'{\i}}z Apell{\'a}niz et al.(2013)]{maiz13} 
Ma{\'{\i}}z Apell{\'a}niz, J., Sota, A., Morrell, N.~I., et al.\ 2013, 
Massive Stars: From alpha to Omega, 198 



\bibitem[{{Ortega} {et~al.}(2013){Ortega}, {Paron}, {Giacani}, {Rubio}, \&
  {Dubner}}]{ortega13}
{Ortega}, M.~E., {Paron}, S., {Giacani}, E., {Rubio}, M., \& {Dubner}, G. 2013,
  \aap, 556, A105

\bibitem[Panagia(1973)]{panagia} Panagia, N.\ 1973, \aj, 78, 
929 


\bibitem[{{Paron} {et~al.}(2011){Paron}, {Petriella}, \& {Ortega}}]{paron11}
{Paron}, S., {Petriella}, A., \& {Ortega}, M.~E. 2011, \aap, 525, A132

\bibitem[Peretto 
\& Fuller(2009)]{peretto09} Peretto, N., \& Fuller, G.~A.\ 2009, \aap, 505, 405 

\bibitem[Pitann et al.(2013)]{pittan13} Pitann, J., Linz, H., 
Ragan, S., et al.\ 2013, \apj, 766, 68 


\bibitem[{{Pound} {et~al.}(2005){Pound}, {Kane}, {Remington}, {Ryutov},
  {Mizuta}, \& {Takabe}}]{pound05}
{Pound}, M.~W., {Kane}, J.~O., {Remington}, B.~A., {et~al.} 2005, \apss, 298,
  177

\bibitem[{{Pound} {et~al.}(2007){Pound}, {Kane}, {Ryutov}, {Remington}, \&
  {Mizuta}}]{pound07}
{Pound}, M.~W., {Kane}, J.~O., {Ryutov}, D.~D., {Remington}, B.~A., \&
  {Mizuta}, A. 2007, \apss, 307, 187

\bibitem[{{Quireza} {et~al.}(2006){Quireza}, {Rood}, {Balser}, \&
  {Bania}}]{quireza06}
{Quireza}, C., {Rood}, R.~T., {Balser}, D.~S., \& {Bania}, T.~M. 2006, \apjs,
  165, 338


\bibitem[{{Rathborne} {et~al.}(2009){Rathborne}, {Johnson}, {Jackson}, {Shah},
  \& {Simon}}]{rath09}
{Rathborne}, J.~M., {Johnson}, A.~M., {Jackson}, J.~M., {Shah}, R.~Y., \&
  {Simon}, R. 2009, \apjs, 182, 131


\bibitem[Reed(2003)]{reed03} Reed, B.~C.\ 2003, \aj, 125, 2531 


\bibitem[{{Rieke} \& {Lebofsky}(1985)}]{rie85}
{Rieke}, G.~H. \& {Lebofsky}, M.~J. 1985, \apj, 288, 618

\bibitem[{{Rieke} {et~al.}(2004){Rieke}, {Young}, {Engelbracht}, {Kelly},
  {Low}, {Haller}, {Beeman}, {Gordon}, {Stansberry}, {Misselt}, {Cadien},
  {Morrison}, {Rivlis}, {Latter}, {Noriega-Crespo}, {Padgett}, {Stapelfeldt},
  {Hines}, {Egami}, {Muzerolle}, {Alonso-Herrero}, {Blaylock}, {Dole}, {Hinz},
  {Le Floc'h}, {Papovich}, {P{\'e}rez-Gonz{\'a}lez}, {Smith}, {Su}, {Bennett},
  {Frayer}, {Henderson}, {Lu}, {Masci}, {Pesenson}, {Rebull}, {Rho}, {Keene},
  {Stolovy}, {Wachter}, {Wheaton}, {Werner}, \& {Richards}}]{rie04}
{Rieke}, G.~H., {Young}, E.~T., {Engelbracht}, C.~W., {et~al.} 2004, \apjs,
  154, 25

\bibitem[{{Roberts} {et~al.}(2011){Roberts}, {van der Tak}, {Fuller}, {Plume},
  \& {Bayet}}]{roberts11}
{Roberts}, H., {van der Tak}, F.~F.~S., {Fuller}, G.~A., {Plume}, R., \&
  {Bayet}, E. 2011, \aap, 525, A107

\bibitem[{{Robitaille} {et~al.}(2007){Robitaille}, {Whitney}, {Indebetouw}, \&
  {Wood}}]{rob07}
{Robitaille}, T.~P., {Whitney}, B.~A., {Indebetouw}, R., \& {Wood}, K. 2007,
  \apjs, 169, 328

\bibitem[Roman-Duval et al.(2009)]{roman09} Roman-Duval, J., 
Jackson, J.~M., Heyer, M., et al.\ 2009, \apj, 699, 1153 




\bibitem[{{Schaerer} \& {de Koter}(1997)}]{sch97}
{Schaerer}, D. \& {de Koter}, A. 1997, \aap, 322, 598

\bibitem[{{Schuller} {et~al.}(2006){Schuller}, {Leurini}, {Hieret}, {Menten},
  {Philipp}, {G{\"u}sten}, {Schilke}, \& {Nyman}}]{schuller06}
{Schuller}, F., {Leurini}, S., {Hieret}, C., {et~al.} 2006, \aap, 454, L87

\bibitem[{{Skrutskie} {et~al.}(2006){Skrutskie}, {Cutri}, {Stiening},
  {Weinberg}, {Schneider}, {Carpenter}, {Beichman}, {Capps}, {Chester},
  {Elias}, {Huchra}, {Liebert}, {Lonsdale}, {Monet}, {Price}, {Seitzer},
  {Jarrett}, {Kirkpatrick}, {Gizis}, {Howard}, {Evans}, {Fowler}, {Fullmer},
  {Hurt}, {Light}, {Kopan}, {Marsh}, {McCallon}, {Tam}, {Van Dyk}, \&
  {Wheelock}}]{skr06}
{Skrutskie}, M.~F., {Cutri}, R.~M., {Stiening}, R., {et~al.} 2006, \aj, 131,
  1163

\bibitem[{{Stern} {et~al.}(2005){Stern}, {Eisenhardt}, {Gorjian}, {Kochanek},
  {Caldwell}, {Eisenstein}, {Brodwin}, {Brown}, {Cool}, {Dey}, {Green},
  {Jannuzi}, {Murray}, {Pahre}, \& {Willner}}]{ste05}
{Stern}, D., {Eisenhardt}, P., {Gorjian}, V., {et~al.} 2005, \apj, 631, 163

\bibitem[{{Stetson}(1987)}]{1987PASP...99..191S}
{Stetson}, P.~B. 1987, \pasp, 99, 191

\bibitem[{{Strom} {et~al.}(1989){Strom}, {Strom}, {Edwards}, {Cabrit}, \&
  {Skrutskie}}]{str89}
{Strom}, K.~M., {Strom}, S.~E., {Edwards}, S., {Cabrit}, S., \& {Skrutskie},
  M.~F. 1989, \aj, 97, 1451

\bibitem[{{Takakuwa} {et~al.}(2007){Takakuwa}, {Ohashi}, {Bourke}, {Hirano},
  {Ho}, {J{\o}rgensen}, {Kuan}, {Wilner}, \& {Yeh}}]{taka07}
{Takakuwa}, S., {Ohashi}, N., {Bourke}, T.~L., {et~al.} 2007, \apj, 662, 431

\bibitem[{{Tenorio-Tagle}(1979)}]{tenorio}
{Tenorio-Tagle}, G. 1979, \aap, 71, 59

\bibitem[Thompson et 
al.(2004a)]{thompson04a} Thompson, M.~A., White, G.~J., Morgan, L.~K., et al.\ 2004, \aap, 414, 1017 

\bibitem[Thompson et
al.(2004b)]{thompson04b} Thompson, M.~A., Urquhart, J.~S., \& White, G.~J.\ 2004, \aap, 415, 627 


\bibitem[{{Tremblin} {et~al.}(2013){Tremblin}, {Minier}, {Schneider}, {Audit},
  {Hill}, {Didelon}, {Peretto}, {Arzoumanian}, {Motte}, {Zavagno}, {Bontemps},
  {Anderson}, {Andr{\'e}}, {Bernard}, {Csengeri}, {Di Francesco}, {Elia},
  {Hennemann}, {K{\"o}nyves}, {Marston}, {Nguyen Luong}, {Rivera-Ingraham},
  {Roussel}, {Sousbie}, {Spinoglio}, {White}, \& {Williams}}]{tremb13}
{Tremblin}, P., {Minier}, V., {Schneider}, N., {et~al.} 2013, \aap, 560, A19

\bibitem[{{van der Tak} {et~al.}(2007){van der Tak}, {Black}, {Sch{\"o}ier},
  {Jansen}, \& {van Dishoeck}}]{vander06}
{van der Tak}, F.~F.~S., {Black}, J.~H., {Sch{\"o}ier}, F.~L., {Jansen}, D.~J.,
  \& {van Dishoeck}, E.~F. 2007, \aap, 468, 627

\bibitem[White et 
al.(1999)]{white99} White, G.~J., Nelson, R.~P., Holland, W.~S., et al.\ 1999, \aap, 342, 233 


\bibitem[{{Werner} {et~al.}(2004){Werner}, {Uchida}, {Sellgren}, {Marengo},
  {Gordon}, {Morris}, {Houck}, \& {Stansberry}}]{wer04}
{Werner}, M.~W., {Uchida}, K.~I., {Sellgren}, K., {et~al.} 2004, \apjs, 154,
  309

\bibitem[{{Wienen} {et~al.}(2012){Wienen}, {Wyrowski}, {Schuller}, {Menten},
  {Walmsley}, {Bronfman}, \& {Motte}}]{wienen12}
{Wienen}, M., {Wyrowski}, F., {Schuller}, F., {et~al.} 2012, \aap, 544, A146

\bibitem[{{Wilson}(1999)}]{wilson99}
{Wilson}, T.~L. 1999, Reports on Progress in Physics, 62, 143

\bibitem[Yamaguchi et al.(1999)]{yama99} Yamaguchi, R., Saito, 
H., Mizuno, N., et al.\ 1999, \pasj, 51, 791 


\bibitem[{{Zacharias} {et~al.}(2010){Zacharias}, {Finch}, {Girard}, {Hambly},
  {Wycoff}, {Zacharias}, {Castillo}, {Corbin}, {DiVittorio}, {Dutta}, {Gaume},
  {Gauss}, {Germain}, {Hall}, {Hartkopf}, {Hsu}, {Holdenried}, {Makarov},
  {Martinez}, {Mason}, {Monet}, {Rafferty}, {Rhodes}, {Siemers}, {Smith},
  {Tilleman}, {Urban}, {Wieder}, {Winter}, \& {Young}}]{zac10}
{Zacharias}, N., {Finch}, C., {Girard}, T., {et~al.} 2010, \aj, 139, 2184

\bibitem[{{Zavagno} {et~al.}(2010){Zavagno}, {Anderson}, {Russeil}, {Morgan},
  {Stringfellow}, {Deharveng}, {Rod{\'o}n}, {Robitaille}, {Mottram},
  {Schuller}, {Testi}, {Billot}, {Molinari}, {di Gorgio}, {Kirk}, {Brunt},
  {Ward-Thompson}, {Traficante}, {Veneziani}, {Faustini}, \&
  {Calzoletti}}]{zav10}
{Zavagno}, A., {Anderson}, L.~D., {Russeil}, D., {et~al.} 2010, \aap, 518, L101


\end{thebibliography}
\end{document}